\documentclass[twocolumn,aps]{revtex4}

\newcommand {\bc}{\begin{center}}
\newcommand {\ec}{\end{center}}
\newcommand {\bea}{\begin{eqnarray}}
\newcommand {\eea}{\end{eqnarray}}
\newcommand {\be}{\begin{equation}}
\newcommand {\ee}{\end{equation}}

\def\lsim{\mathrel{\rlap{\lower4pt\hbox{\hskip1pt$\sim$}}
    \raise1pt\hbox{$<$}}}
\def\gsim{\mathrel{\rlap{\lower4pt\hbox{\hskip1pt$\sim$}}
    \raise1pt\hbox{$>$}}}
\usepackage{graphicx}
\usepackage{color}
\usepackage{soul}

\begin{document}


\title{Determination of the density and temperature dependence 
of the shear viscosity of a unitary Fermi gas based on 
hydrodynamic flow}

\author{Marcus~Bluhm$^{1,2}$, Jiaxun~Hou$^1$, Thomas~Sch\"afer$^1$}

\affiliation{$^1$Department of Physics, North Carolina State University,
Raleigh, NC 27695, USA\\
$^2$Institute of Theoretical Physics, University of Wroc\l{}aw,
50204 Wroc\l{}aw, Poland}

\begin{abstract}
We determine the shear viscosity of the ultracold Fermi gas at 
unitarity in the normal phase using hydrodynamic expansion data. The 
analysis is based on a generalized fluid dynamic framework which ensures
a smooth transition between the fluid dynamic core of the cloud
and the ballistic corona. We use expansion data taken by Joseph et 
al.~\cite{Joseph:2014} and measurements of the equation of state
by Ku et al.~\cite{Ku:2011}. We find that the shear viscosity to 
particle density ratio just above the critical temperature is
$\left.\eta/n\right|_{T_c}=0.41\pm 0.11$. We also obtain evidence
that the shear viscosity to entropy density ratio has a minimum
slightly above $T_c$ with $\left.\eta/s\right|_{\it min}=0.50\pm 0.10$.

\end{abstract}

\maketitle

{\it Introduction:}
 The dilute Fermi gas at unitarity is a very attractive physical system 
for studying the transport properties of strongly correlated quantum 
fluids \cite{Schafer:2009dj,Adams:2012th,Schaefer:2014awa}. From a 
theoretical point of view the unitary Fermi gas is a parameter-free,
scale invariant, and intrinsically quantum mechanical many-body system. 
A lot of interest has centered on the question of how close the viscosity
to entropy density ratio of this system comes to the proposed string
theory bound $\eta/s=\hbar/(4\pi k_B)$ \cite{Kovtun:2004de}.
Experimentally, the unitary Fermi gas can be realized in dilute 
atomic gases using Feshbach resonances \cite{Bloch:2007,Giorgini:2008}. 
The experimental control provided by Feshbach resonances implies that
we can study the transition from the strongly correlated unitary
Fermi gas to weakly coupled Bose and Fermi gases. 

 In this work we focus on the problem of extracting the shear viscosity
of the unitary Fermi gas from experiments with trapped ultracold gases
\cite{Kinast:2004b,Schafer:2007pr,Turlapov:2007,Bruun:2007,Cao:2010wa,Elliott:2013b,Joseph:2014,Elliott:2013,Brewer:2015hua}.
Our main interest is in the low temperature regime, where the density 
dependence of the shear viscosity is relevant, and the minimum of $\eta/s$
is likely to be achieved. There are two main types of experiments that are 
relevant to this problem. The first class involves measuring the damping 
rate of collective excitations, and the second focuses on the expansion of 
the cloud after removing the trapping potential. From a theoretical 
perspective the damping experiments would appear to be more attractive, 
because even a very small viscosity leads to a clear signature in the 
exponential decay of the collective mode. In practice, however, the expansion 
experiments take place in a cleaner environment and have achieved greater 
accuracy. In an expansion experiment what is observed is the time evolution 
of the aspect ratio of the cloud. Hydrodynamic pressure gradients accelerate 
the cloud along the short direction, so that the aspect ratio increases as 
a function of time. Viscosity counteracts the pressure gradients, and slows 
the growth of the aspect ratio. These flow experiments are very similar
to elliptic flow experiments in relativistic heavy ion physics
\cite{Ackermann:2000tr,Adler:2003kt,Aamodt:2010pa}.

 The main difficulty in analyzing these experiments is that the viscosity
$\eta(n,T)$ is a local quantity that varies with the density $n$ and 
temperature $T$ of the cloud, while the observed aspect ratio is a global 
property of the trapped gas. This means that the dependence of the 
data on initial cloud energy, particle number, and expansion time has
to be unfolded to determine $\eta(n,T)$. An even more significant 
problem is that the viscosity is a parameter that appears in the 
fluid dynamic description of the cloud. However, fluid dynamics breaks
down in the dilute, dissipative corona of the gas. 

 We have recently made significant progress in dealing with the physics
of the dilute corona. We have introduced a new method, anisotropic 
fluid dynamics \cite{Bluhm:2015raa,Florkowski:2010,Martinez:2010sc}, 
that takes into account the effects of non-hydrodynamic modes. These modes 
quickly relax in the dense part of the cloud so that Navier-Stokes fluid
dynamics is recovered. In the dilute corona non-hydrodynamic modes
ensure a smooth transition to a free-streaming, ballistic expansion. 
We have checked numerically that anisotropic fluid dynamics reproduces
the Navier-Stokes equation in the dense limit \cite{Bluhm:2015raa} as 
well as numerical solutions of the Boltzmann equation in the dilute 
regime \cite{Pantel:2014jfa,Bluhm:2015bzi}. We have also shown that 
the anisotropic fluid dynamics, combined with the kinetic theory
prediction for the shear viscosity $\eta=15/(32\sqrt{\pi})(mT)^{3/2}$
\cite{Bruun:2005}, reproduces the high temperature expansion data
obtained in \cite{Cao:2010wa}. Note that here and in the remainder
of the paper we set $\hbar$ and $k_B$ equal to unity. 

 In this work we extend our studies to lower temperature. For this 
purpose we fit the expansion data to a systematic expansion of the
viscosity in powers of the density. We show that the data clearly demand 
that the shear viscosity has non-trivial density dependence. We also show 
that the density dependence in the normal phase is quite smooth, and that 
the existing data place strong constraints on $\eta/n$ near $T_c$. This 
study requires several refinements of our previous work. We extend the 
fluid dynamic analysis to three dimensional systems with no axial symmetry. 
We include an accurate parametrization of the measured equation of state, 
and a more general functional form of the shear viscosity. 

{\it Anisotropic fluid dynamics:}
 In this section we briefly summarize the anisotropic fluid dynamics
method \cite{Bluhm:2015raa}. The fluid dynamical variables 
that characterize a non-relativistic fluid in the normal phase are the 
mass density $\rho$, the momentum density $\vec{\pi}=\rho\vec{u}$, and 
the energy density ${\cal E}$. The equations of motion follow from the 
conservation laws 
\bea
\label{rho_lag}
 D_0\rho &=& -\rho \,\vec{\nabla}\cdot\vec{u}\, ,  \\
\label{u_lag}
 D_0 u_i & = & - \frac{1}{\rho}\, \nabla_j \left( \delta_{ij} P 
 + \delta \Pi_{ij} \right) \, , \\
\label{e_lag}
 D_0 \epsilon & = & - \frac{1}{\rho}\, \nabla_i 
  \left( u_i P + \delta \jmath^{\cal E}_i \right) \, . 
\eea
Here, we defined the comoving time derivative $D_0=\partial_0 +\vec{u}
\cdot\vec{\nabla}$, the energy per mass $\epsilon={\cal E}/\rho$, and the 
pressure $P$. We also introduce the energy density in the rest frame
of the fluid, ${\cal E}^0={\cal E}-\frac{1}{2}\rho\vec{u}^{\, 2}$. In 
order for the equations to close we have to provide an equation of state 
$P=P({\cal E}^0,\rho)$, and constitutive equations for the dissipative 
stresses $\delta\Pi_{ij}$ and the dissipative energy current $\delta
\jmath^{\cal E}_i$. For the unitary Fermi gas scale invariance implies 
that $P=\frac{2}{3}{\cal E}^0$.

 In the Navier-Stokes approximation the dissipative stresses are expanded
to first order in gradients of the thermodynamic variables. We get $\delta
\Pi_{ij}=-\eta\sigma_{ij}$ with 
\be 
\sigma_{ij} =\nabla_i u_j + \nabla_j u_i -\frac{2}{3}\delta_{ij} 
  \vec\nabla\cdot\vec{u}
\ee
and $\delta \jmath^{\cal E}_i = u_j\delta \Pi_{ij}$. Scale invariance implies
that the bulk viscosity vanishes. We have also used the fact that in 
expansion experiments the effects of heat conduction are of higher order 
in the gradient expansion. This is related to the fact that the initial 
temperature is constant, and that the expansion of an ideal gas preserves 
the isothermal nature of the temperature profile \cite{Schafer:2010dv}.

 In anisotropic fluid dynamics we treat the components of the dissipative 
stress tensor as independent fluid dynamical variables. The symmetries of
the trap imply that the stresses are diagonal. We define anisotropic 
components of the pressure, $P_a$ for $a=1,2,3$, and define 
\be
\delta\Pi_{ij}={\rm diag}(\Delta P_1,\Delta P_2,\Delta P_3), 
\ee
where $\Delta P_a=P_a-P$. We also define anisotropic components of the 
energy density ${\cal E}_a$ such that ${\cal E}=\sum_a{\cal E}_a$. The 
anisotropic components of the energy per mass satisfy the equation of 
motion \cite{Bluhm:2015raa}
\be 
\label{e_a_lag}
 D_0 \epsilon_a = - \frac{1}{\rho}
 \nabla_i \left[ \delta_{ia} u_i P + (\delta \jmath^{\cal E}_a)_i \right] 
 - \frac{P}{2\eta\rho} \,\Delta P_a\, , 
\ee
where $\epsilon_a={\cal E}_a/\rho$ and $(\delta \jmath^{\cal E}_a)_i 
= \delta_{ia} u_j \delta \Pi_{ij}$. The anisotropic pressure is related
to the anisotropic energy density by an equation of state. In the case of 
a scale invariant fluid we have $P_a({\cal E}^0_a)=2\,{\cal E}^0_a$ with 
${\cal E}^0_a={\cal E}_a-\frac{1}{2}\rho u_a^2$. Then $P=\frac{1}{3}\sum_a 
P_a$ satisfies the isotropic equation of state, and equ.~(\ref{e_a_lag}) 
gives the isotropic equation of energy conservation equ.~(\ref{e_lag}) when 
summed over $a$. In our previous work we have described a three dimensional 
fluid dynamics code that solves equ.~(\ref{rho_lag})-(\ref{e_lag}) and 
equ.~(\ref{e_a_lag}) \cite{Schafer:2010dv,Bluhm:2015raa}. This code is 
based on the PPM scheme of Colella and Woodward 
\cite{Colella:1984,Blondin:1993}.

 We have shown that in the limit of small viscosity, $\eta
(\vec{\nabla}\cdot\vec{u})\ll P$, the anisotropic pressure terms 
relax to the viscous stress tensor in Navier Stokes theory, $\Delta P_a
=-\eta\sigma_{aa}$. We observe that in the opposite limit, that of very 
large viscosity, equ.~(\ref{e_a_lag}) becomes a conservation law. This
conservation law ensures that anisotropic fluid dynamics reproduces the 
free streaming limit. Finally, we have checked that anisotropic fluid
dynamics provides a very accurate representation of numerical solutions
of the Boltzmann equation in the limit that two-body scattering 
dominates \cite{Bluhm:2015bzi}.

 In general the viscosity is a function of density and temperature. 
In the unitary limit scale invariance implies that $\eta(n,T)= (mT)^{3/2}
f(n\lambda^3)$, where $\lambda=[(2\pi)/(mT)]^{1/2}$ is the de Broglie
wave length. In this work we will expand the function $f(x)$ in powers
of the diluteness of the gas
\be
 \eta(n,T)=\eta_0(mT)^{3/2} \left\{ 1 + \eta_2 \left(n\lambda^3\right)
 + \eta_3 \left( n\lambda^3\right)^2 + \ldots \right\} \, . 
\ee
We note that the leading term is purely a function of temperature, 
the first correction is solely a function of density, and higher 
order terms depend on increasing powers of the density. In general
this expansion is not expected to be useful near $T_c$, but we will
show that terms that scale as $(n\lambda^3)^2$ and higher are
surprisingly small. 

{\it Experimental parameters:}
 We will analyze the expansion data reported in~\cite{Joseph:2014}. 
This work represents the most complete set of elliptic flow measurements
for the unitary Fermi gas over a wide range of temperatures currently 
available. The gas is released from a harmonic trap $V_{\it ext}=\frac{1}{2}
m\omega_i^2x_i^2$ with trap frequencies $(\omega_x,\omega_y,\omega_z)=
(2\pi)(2210,830,64.3)$ {\it Hz}. After the optical trap is turned off
there is a residual magnetic bowl characterized by $\omega_{\it mag}=2\pi
\cdot 21.5\,{\it Hz}$. The total energy per particle of the gas varies between
$E/(NE_F)=(0.56-1.91)$. Here, $N$ is the number of particles and $E_F
\equiv (3N)^{1/3}\bar\omega$, where $\bar\omega$ is the geometric mean
of the trap frequencies.  The energy and temperature of the cloud are 
extracted using absorption images and an equation of state ${\cal E}^0
(n,T)$. We describe a parametrization of the equation of state 
measured by the MIT group \cite{Ku:2011} in the Appendix. 
Based on this equation of state we find that the critical energy where 
superfluidity occurs at the center of the trap is $E/(NE_F)=0.70$. In 
the high temperature limit many relations simplify. For example, the 
total cloud energy is given by $E=3NT$. We will characterize the initial 
temperature using the dimensionless ratio $T/T_F$, where $T_F=E_F$. 

\begin{figure}[t]
\bc\includegraphics[width=7.5cm]{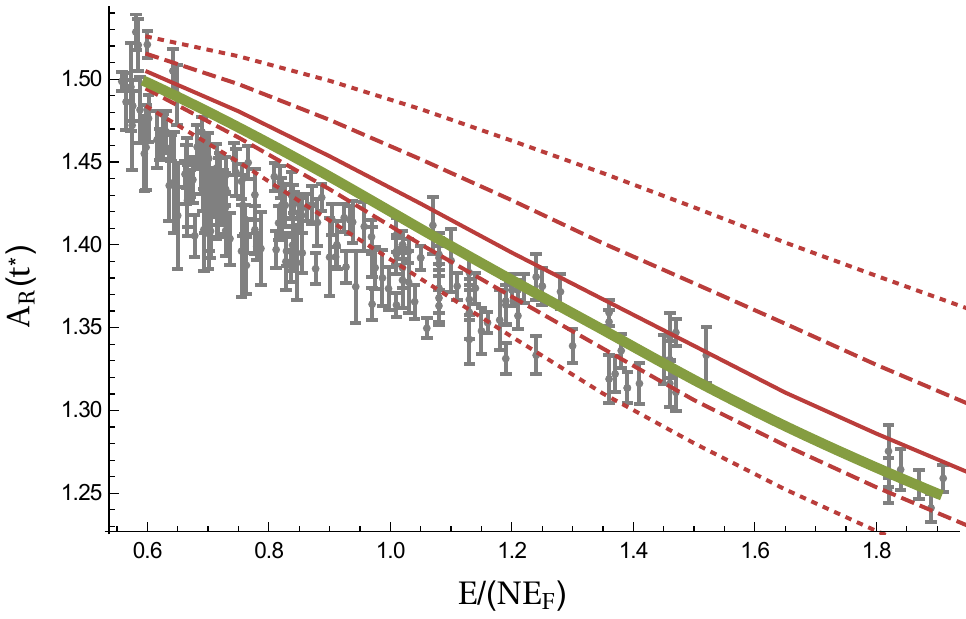}\ec
\caption{\label{fig_Ar_E0_fit}
Aspect ratio $A_R=\sigma_x/\sigma_y$ at $t^*=1.2\,{\it msec}$ as a function
of the energy $E/(NE_F)$ of the cloud. Data (gray points) compared to 
hydrodynamic fits based 
on the equation of state of a free gas. The solid red line corresponds to 
the shear viscosity $\eta=\eta_0(mT)^{3/2}$ predicted by kinetic theory, and 
the dashed and dotted line show the $\pm 25\%$ and $\pm 50\%$ range in 
$\eta_0$. The thick green line is the best fit to the high energy data, 
corresponding to $\eta_0=0.301$.}
\end{figure}

\begin{figure}[t]
\bc\includegraphics[width=7.5cm]{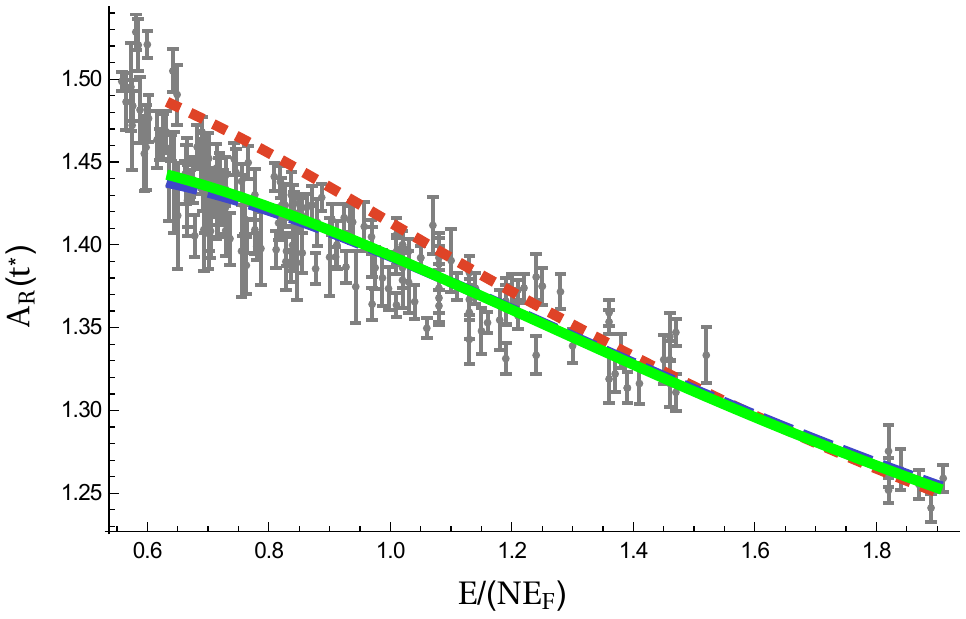}\ec
\caption{\label{fig_Ar_E0_EOS_vir}
Aspect ratio $A_R=\sigma_x/\sigma_y$ at $t^*=1.2\,{\it msec}$ as a function
of the energy $E/(NE_F)$ of the cloud. Data (gray points) compared to 
hydrodynamic fits based on the measured equation of state. The red short-dashed 
line shows the high temperature fit, the blue dashed line includes density 
corrections, and the green solid line contains a density squared term.}
\end{figure}

{\it Scaling of the aspect ratio with the initial energy:}
 Expansion experiments measure the time evolution of the aspect
ratio $A_R(t)$ for different initial energies and particle 
numbers. The experiment of Joseph et al.~\cite{Joseph:2014} focuses
on the ratio $\sigma_x/\sigma_y$, which reaches its asymptotic
behavior more quickly than $\sigma_x/\sigma_z$ or $\sigma_y/\sigma_z$.
The radii $\sigma_i$ are determined from a Gaussian fit to 
two-dimensional absorption images. As noted in \cite{Pantel:2014jfa}
it is important to follow this definition when analyzing the data 
using transport theory. In particular, there is a significant 
difference between the ratio of rms radii, $\sqrt{\langle x^2\rangle}/
\sqrt{\langle y^2\rangle}$, and the ratio of Gaussian fit radii, 
$\sigma_x/\sigma_y$. This is the case even if the initial density 
distribution is a Gaussian. 

 Joseph et al.~observed that the main information about the density
and temperature dependence of $\eta(n,T)$ is not carried by the time 
dependence of $A_R(t)$ for fixed initial energy, but by the dependence 
of $A_R(t^*)$ at a fixed time $t^*$ on the initial energy. In 
Fig.~\ref{fig_Ar_E0_fit} we show $A_R(t^*)=\sigma_x/\sigma_y$ as a 
function of $E/(NE_F)$ at $t^*=1.2\,{\it msec}$. Note that the plot
covers a fairly narrow range in $A_R$. Individual data points are more 
accurate than previously published data, which spanned a much larger 
range in aspect ratio. 

 A difficulty in interpreting the results is that the data points 
correspond to a range of particle numbers. The data are 
clustered around a mean $\bar{N}=1.94\cdot 10^5$, and the variance 
in $N^{1/3}$, which is relevant to the effective viscosity, is about 7\%. 
We show all the data points on the same plot, but when performing 
hydrodynamic fits we use the correct number of particles for each 
individual data point.

Figure~\ref{fig_Ar_E0_fit} shows a fit to the data based on the high 
temperature theory only. This means that we use the free gas equation 
of state, and only the first coefficient, $\eta_0$, in the virial 
expansion of the shear viscosity. The best fit to the high temperature 
data gives $\eta_0=0.301$
which is somewhat higher than the value $\eta_0=0.264$ 
predicted by kinetic theory. The best fit value shifts slightly if the 
full equation of state is used, but the shape of $A_R(t^*)$ as a function 
of $E/(NE_F)$ does not change. We observe that the data at lower energy 
clearly demand a more complicated functional form of the shear viscosity. 

\begin{figure}[t]
\bc\includegraphics[width=7.5cm]{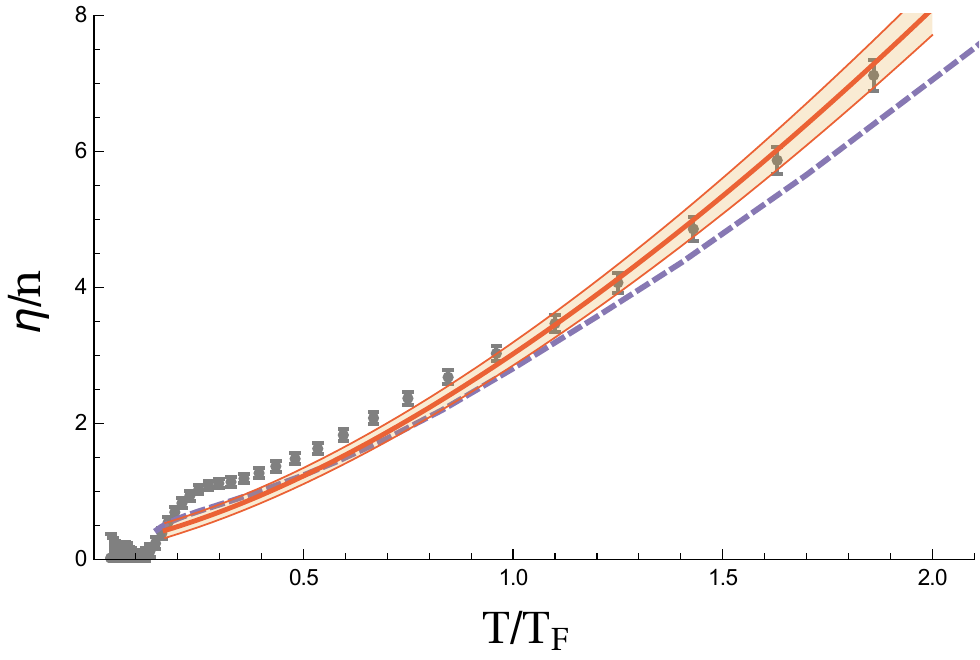}\ec
\caption{\label{fig_alpha_n}
In this figure we show the reconstructed ratio $\eta/n$ 
as a function of $T/T_F^{\it loc}$ for a homogeneous gas. The local 
Fermi temperature is defined as $T_F^{\it loc}=k_F^2/(2m)$ where $k_F$
is defined via the density of the gas, $n=k_F^3/(3\pi^2)$. The
red line shows the density expansion together with the error band
described in the text. The curves terminate at $T_c$. The gray
dots show the reconstruction obtained in \cite{Joseph:2014}, and 
the dashed line shows the $T$-matrix calculation of Enss et 
al.~\cite{Enss:2010qh}.}
\end{figure}

 Figure~\ref{fig_Ar_E0_EOS_vir} shows a fit to the data above the superfluid
transition based on the full equation of state and an expansion of the 
shear viscosity up to second order in density.
The best fit is 
\be 
\eta_0 = 0.265\pm 0.02\, ,\hspace{1cm}
\eta_2 = 0.060\pm 0.02\, ,
\ee
and $\eta_3=-(2\pm 8)\cdot 10^{-4}$. We observe that the $n^2$ coefficient 
is consistent with zero within error bars. We also find that the fit is 
stable with respect to including higher order terms in $n$. The 
$\chi^2/N_{\it dof}$ of the fit is of order unity, indicating that 
this simple model provides a very good representation of all the data 
in the entire regime above the superfluid phase transition. We note 
that $\eta_0$ agrees to better than 1\% with the kinetic theory prediction 
$\eta_0=0.264$.

\begin{figure}[t]
\bc\includegraphics[width=7.5cm]{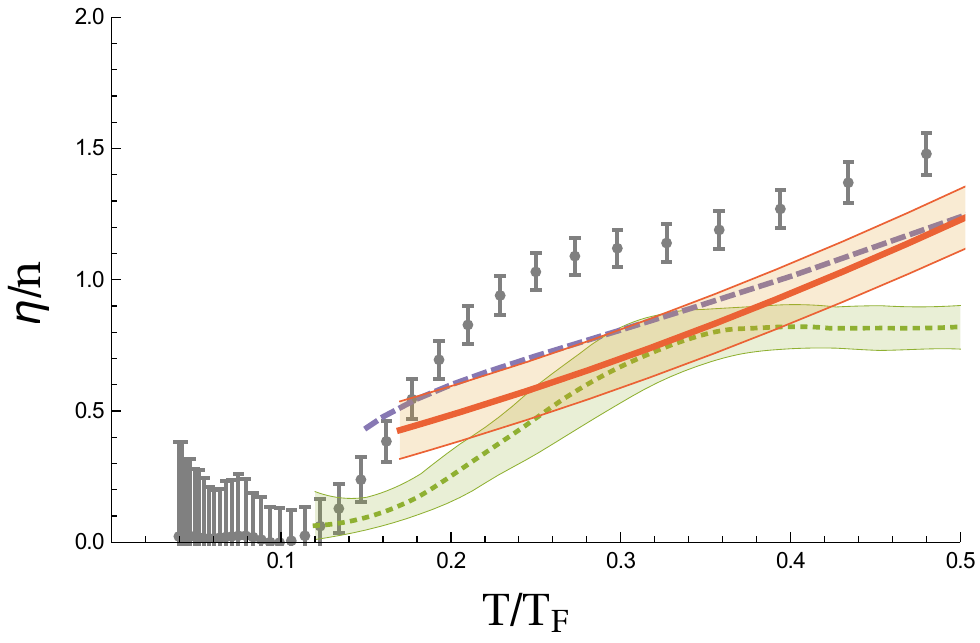}\ec
\caption{\label{fig_alpha_n_zoom}
Same as Fig.~\ref{fig_alpha_n}, zooming in on the low temperature
regime. Our analysis (red band) is compared to the results (gray 
points) obtained in \cite{Joseph:2014}, the $T$-matrix calculation 
(dashed line) of Enss et al.~\cite{Enss:2010qh}, and the lattice
calculation (green band) of Wlazlowski et al.~\cite{Wlazlowski:2013owa}.}
\end{figure}

{\it Conclusions:}
 Our determination of $\eta/n$ for the homogeneous Fermi gas is shown in 
Figs.~\ref{fig_alpha_n} and \ref{fig_alpha_n_zoom}. The result is shown as a 
function of $T/T_F^{\it loc}$, where $T_F^{\it loc}=k_F^2/(2m)$ is the local Fermi 
temperature of the gas. The best fit to the data, shown as the thick red line 
in Fig.~\ref{fig_alpha_n}, is 
\be
\label{eta_n_fit}
\eta/n=2.773\,x^{3/2} + 0.251 -0.0013\,x^{-3/2}\, , 
\ee
where $x=T/T_F^{\it loc}$. The coefficients in equ.~(\ref{eta_n_fit})
are given by the central values of $\eta_{0},\eta_1,\eta_2$ normalized
by the density $n$. The local Fermi momentum $k_F$ is defined in terms 
of the density of the gas, $n=k_F^3/(3\pi^2)$. We show the reconstruction
for temperatures above the critical temperature $T_c =0.167(13)T_F^{\it loc}$ 
\cite{Ku:2011}. We find that the value of the viscosity at $T_c$ is 
$\eta/n|_{T_c}=0.41\pm 0.11$. We have not attempted to reconstruct the 
shear viscosity below $T_c$, since a proper treatment of this regime requires 
superfluid hydrodynamics.

\begin{figure}[t]
\bc\includegraphics[width=7.5cm]{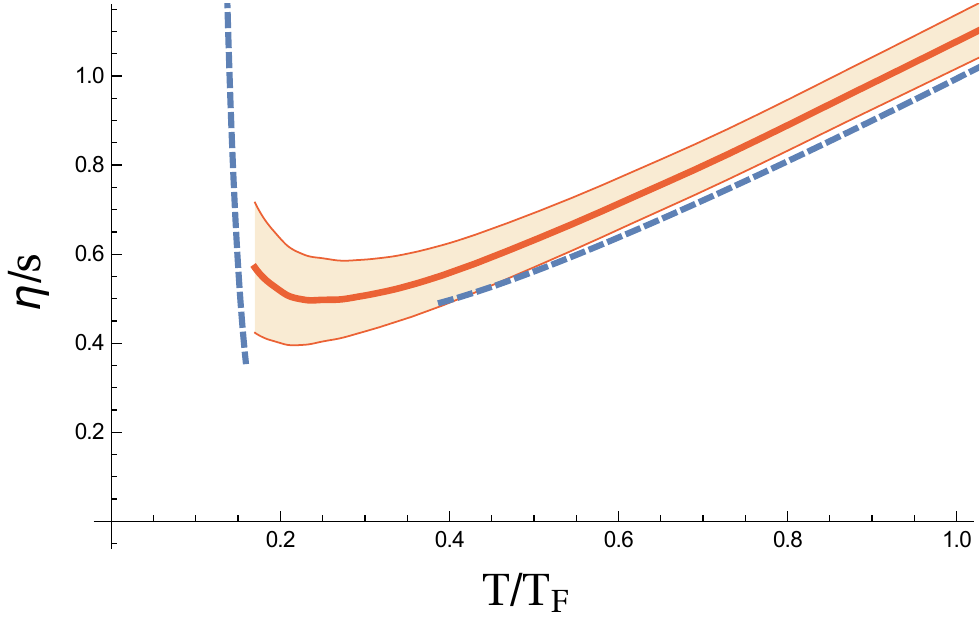}\ec
\caption{\label{fig_alpha_s}
Low temperature behavior of the shear viscosity to entropy density ratio
$\eta/s$ as a function of $T/T_F^{\it loc}$. Our analysis (red band) is compared
to the high and low temperature predictions from kinetic theory, see
\cite{Bruun:2005,Rupak:2007vp}.}
\end{figure}

 For comparison the gray data points show the reconstructed values
of $\eta/n$ obtained in the experimental work of Joseph et 
al.~\cite{Joseph:2014}. These results are based on the same expansion data, but 
involve a number of assumptions \footnote{The $\chi^2/N_{\it dof}$ of our
fit is close to one, whereas the reconstruction of \cite{Joseph:2014}, 
used as an input to our hydrodynamic model, has $\chi^2/N_{\it dof}\simeq
4.5$. More importantly, we stress that the result in \cite{Joseph:2014}
has an unknown overall normalization, which was fixed using kinetic theory.}. 
The main assumption is that there 
is a critical radius $R^{\it crit}_i$ so that the atomic cloud inside
this radius can be described as a viscous fluid, and the particles
outside the radius are a non-interacting gas. The critical radius 
is assumed to be a constant fraction of the cloud size.  
The overall constant is adjusted to reproduce the 
expected behavior of the high temperature viscosity, $\eta\sim\eta_0
(mT)^{3/2}$. This implies that the agreement of the data points with
kinetic theory for large $T/T_F^{\it loc}$ is not a result, but an input. 
In contrast, the agreement of our reconstruction with kinetic theory 
is a non-trivial result. There is some discrepancy between the two 
reconstructions in the regime $T=(0.2-1.0)T_F^{\it loc}$. In this regime
our result for $\eta/n$ is systematically lower. This makes sense
if one assumes that as the temperature is lowered and the viscosity
drops the effective fluid radius increases. This implies that 
assuming a constant radius of the fluid core leads to an overestimate
of the viscosity. It is interesting that directly at $T_c$ the 
two reconstructions agree. 
 
 We also show the $T$-matrix calculation of Enss et al.~\cite{Enss:2010qh}, 
which agrees quite well with our reconstructed viscosity near $T_c$. It 
will be interesting to study the physical consequences of this result, 
for example possible implications for quasi-particle models. We also 
show the lattice calculation of Wlazlowski et al.~\cite{Wlazlowski:2013owa}.
The calculation does not match the shape of our reconstruction, and has
a substantially smaller $\left.\eta/n\right|_{T_c}$.

 Finally, Fig.~\ref{fig_alpha_s} shows the ratio of shear viscosity 
to entropy density, based on our reconstruction of $\eta/n$ and the 
measurement of $s/n$ by the MIT group \cite{Ku:2011}. The result is 
compared to high and low temperature predictions for $\eta/s$ in 
kinetic theory \cite{Bruun:2005,Rupak:2007vp}. We find a shallow 
minimum of $\left.\eta/s\right|_{\it min}=0.50\pm 0.10$ slightly 
above $T_c$. The minimum is related to the fact that the entropy
per particle drops siginificantly as $T_c$ is approached from above, 
whereas no structure is seen in $\eta/n$. We note that at present we 
can only weakly exclude (at about 1$\sigma$) a minimum in $\eta/s$
at or below $T_c$. A minimum in $\eta/s$ above $T_c$ was predicted 
in \cite{Zwerger:2016xma}, but is in tension with the Monte Carlo
data in \cite{Wlazlowski:2013owa}.

Acknowledgments: This work was supported in parts by the US Department 
of Energy grant DE-FG02-03ER41260. The work of M.~Bluhm is funded by the 
European Union's Horizon~2020 research and innovation program under the 
Marie Sk\l{}odowska Curie grant agreement No 665778 via the National 
Science Center, Poland, under grant Polonez UMO-2016/21/P/ST2/04035.
We would like to thank James Joseph 
and John Thomas for many useful discussions, and for providing us with
the data in \cite{Joseph:2014}. We thank Martin Zwierlein for providing
us with the data in \cite{Ku:2011}.

\begin{appendix}
\section{Choice of units and scaling with the number of particles}
\label{sec_vir}

 The fluid dynamic analysis employs a set of dimensionless variables. 
The time scale for the expansion is set by the inverse geometric mean of 
the trap frequencies, $t_0={\bar\omega}^{-1}= 3.24534\cdot10^{-4}\,{\it sec}$, 
and a dimensionless time variable is given by $\bar{t}=t/t_0$. We will also 
use dimensionless frequency variables $\tilde\omega_i=\omega_i/\bar{\omega}$. 
In the present case $(\tilde\omega_x,\tilde\omega_y,\tilde\omega_z) = 
(4.506, 1.692, 0.131)$. The unit of distance is 
\be 
 x_0 = \left[ \frac{2}{3}\frac{(3N)^{1/3}}{m\bar\omega}\right]^{1/2}\, .
\ee
For a typical number of atoms, $N=2\cdot 10^5$, this is $x_0 = 13.8\,\mu 
{\it m}$. The unit of density is $n_0=x_0^{-3}$, the unit of temperature 
is $T_0=m\bar{\omega}^2x_0^2$, and the unit of pressure is $P_0=n_0T_0=
m\bar{\omega}^2x_0^{-1}$. The unit of viscosity is $\eta_0=m\bar\omega/x_0$. 
We can construct dimensionless variables by considering appropriate ratios. 
The dimensionless temperature is $\bar{T}=T/T_0$. In the case of densities 
we also normalize by the density $n^B$ of a Boltzmann gas at the center 
of the trap. We have $\bar{n} = n/n^{\it B}(0)$, $\bar{P} = (P/P_0)(n_0/
n^{\it B}(0))$ and $\bar{\eta} = (\eta/\eta_0)(n_0/n^{\it B}(0))$
with 
\be 
\frac{n^{\it B}(0)}{n_0} =  \frac{N}{(3\pi)^{3/2}} 
                  \left(\frac{T_F}{T}\right)^{3/2}\, . 
\ee
These rescalings preserve simple relations such as the equation
of state of a free gas $P=nT$, which becomes $\bar{P}=\bar{n}\bar{T}$.
In scaled variables the initial condition in the high temperature limit
takes a very simple form. We have 
\be 
\bar{n}(\bar{x})= \exp[ -(\tilde\omega_i^2\bar{x}_i^2)/(2\bar{T})],
\ee
which is independent of the number of particles $N$. Initial density and 
pressure profiles that follow from the full equation of state are discussed in 
Sec.~\ref{sec_eos} below. The exact profiles are also independent of $N$. 

 The dependence on the number of particles enters through viscous
corrections. The density expansion of the viscosity can be written
as
\be
\eta=\alpha_T (mT)^{3/2} + \alpha_{n}n + \alpha_{n^2}n^2(mT)^{-3/2}+ \ldots,
\ee
where the coefficients $\alpha_i$ are related to $\eta_i$ by 
\bea
\alpha_T     &=&  \eta_0 \, , \\
\alpha_n     &=&  (2\pi)^{3/2}\eta_0\eta_2\, , \\
 \alpha_{n^2} &=&  (2\pi)^3\eta_0\eta_3  \, . 
\eea
In dimensionless variables we write $\bar\eta = \bar\alpha_T \bar{T}^{3/2} + 
\bar\alpha_n\bar{n} + \bar\alpha_{n^2}\bar{T}^{-3/2}\bar{n}^2 + \ldots$,
where 
\bea
\label{alpha_T}
\bar\alpha_T   &=& \frac{4\pi^{3/2} \alpha_T}{3(3N)^{1/3}}
                      \left(\frac{3T}{T_F}\right)^{3/2}\, ,  \\
\bar{\alpha}_n &=& \frac{3\alpha_n}{2(3N)^{1/3}} \, , \\
\bar\alpha_{n^2}&=& \frac{27\alpha_{n^2}}{16\pi^{3/2}(3N)^{1/3}}
                      \left(\frac{T_F}{3T}\right)^{3/2} \, . 
\eea 
We observe that at fixed $T/T_F$ all terms in the density 
expansion of the viscosity scale as $N^{-1/3}$. This makes it
relatively easy to analyze data taken at different values 
of $N$. 

 In the manuscript we provide a simple fit to the viscosity which 
is of the form $\eta/n= a_0 x^{3/2} + a_1 + a_2 x^{-3/2}$, where
$x=T/T_F^{\it loc}$ with $T_F^{\it loc}=k_F^2/(2m)$ and 
$n=k_F^3/(3\pi^2)$. This result is useful for comparing the fluid dynamic
fits with theoretical and numerical results in the literature.
The coefficients $a_i$ are given by 
\bea
a_0 &=& \alpha_T \, \frac{3\pi^2}{2^{3/2}}\, , \\
a_1 &=& \alpha_n \, , \\
a_2 &=& \alpha_{n^2}\, \frac{2^{3/2}}{3\pi^2} \, . 
\eea

\section{Equation of state}
\label{sec_eos}

\begin{figure}[t]
\bc\includegraphics[width=7.5cm]{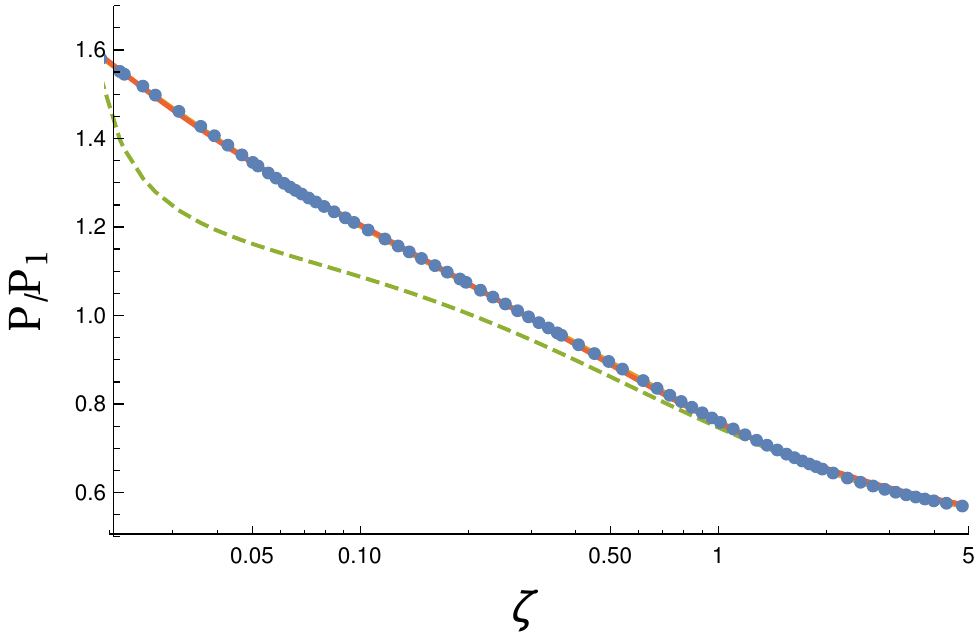}\ec
\caption{\label{fig_EOS_h}
Ratio of the full pressure over the pressure of a free gas, as a function 
of inverse fugacity, from Fig.~4 of \cite{Ku:2011}. The points show the data 
points, and the line is the fit function discussed in the text. The dashed 
line shows a fit to the data of Nascimbene et al.~\cite{Nascimbene:2009}.
The critical point corresponds to $\zeta\simeq 0.08$.}
\end{figure}

 The high temperature analysis in \cite{Bluhm:2015bzi} was based
on the non-interacting equation of state $P=nT$. In order to perform 
a reliable determination of $\eta(n,T)$ at lower temperature we need 
to use a realistic equation of state. It is interesting to note that 
the viscous fluid dynamics evolution does not make use of $P(n,T)$, 
it only uses the exact relation $P=\frac{2}{3}{\cal E}$. However, 
the equation of state is needed to convert the experimentally measured
quantity $E/(NE_F)$ to an initial temperature $T/T_F$, and to determine 
the initial density and pressure profile $n(x,0)$, $P(x,0)$. The 
equation of state is also needed to determine the local temperature 
that enters into the shear viscosity $\eta(n,T)$. 

 In the following we will discuss a simple parametrization of the 
equation of state measured by the MIT group \cite{Ku:2011}. The 
method we have used is described in Appendix A of \cite{Schafer:2010dv},
where it was used to parameterize the equation of state published
by Nascimbene et al.~\cite{Nascimbene:2009}. A similar fit was also
employed in \cite{Romatschke:2012sf}. We write the pressure as 
\be 
\label{app_1}
 P(\zeta)= h(\zeta)P_1(\zeta), \hspace{0.5cm}
 P_1(\zeta) =-T\lambda^{-3} {\it Li}_{5/2}(-\zeta^{-1})\, , 
\ee
where $\zeta=\exp(-\mu/T)$ is the inverse fugacity, $P_1$ is the pressure
of a single component non-interacting Fermi gas, $\lambda=[(2\pi)/(mT)]^{1/2}$ 
is the thermal de Broglie wave length, and ${\it Li}_p$ is the Polylogarithm 
of order $p$. We also define 
\be
f(\zeta)=-h(\zeta){\it Li}_{5/2}(-\zeta^{-1})\, .
\ee
Numerical data for $h(\zeta)$ are shown in Fig.~\ref{fig_EOS_h}.
The data are well fit by 
\be
\label{app_2}
\frac{h(\zeta)}{2} = 
  \frac{\zeta^2+c_1\zeta+c_2}{\zeta^2+c_3\zeta+c_4}
\ee
with 
\bea 
\label{app_3}
c_1=1.32109,&\hspace{0.2cm} & 
c_2=0.026341,\\
c_3=0.541993,&\hspace{0.2cm}& 
c_4=0.005660. \nonumber
\eea
If we expand the function $h(\zeta)$ we get $h(\zeta)/2\simeq
1 + 0.779/\zeta$, consistent with the theoretical value for the
second virial coefficient, $b_2=1/\sqrt{2}\simeq 0.71$. 

\begin{figure}[t]
\bc\includegraphics[width=7.5cm]{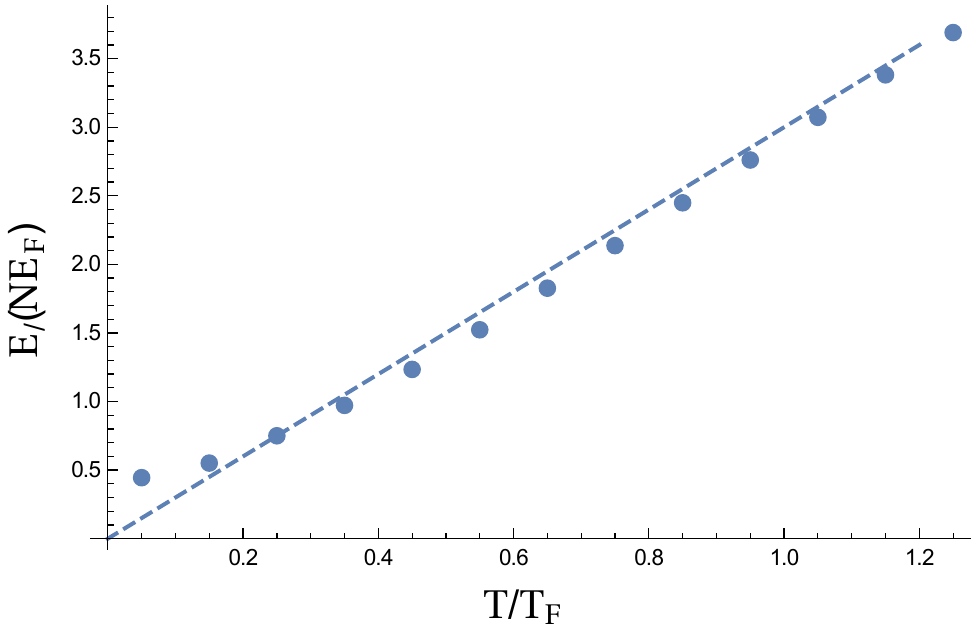}\ec
\caption{\label{fig_EoEf}
Relation between the energy of the trapped gas $E/(NE_F)$ and the 
temperature of the cloud $T/T_F$. The data points are computed using 
a parameterization of the MIT equation of state \cite{Ku:2011}, and the 
dashed line shows the high temperature limit $E/(NE_F)=3(T/T_F)$.}
\end{figure}

\begin{figure}[t]
\bc\includegraphics[width=7.5cm]{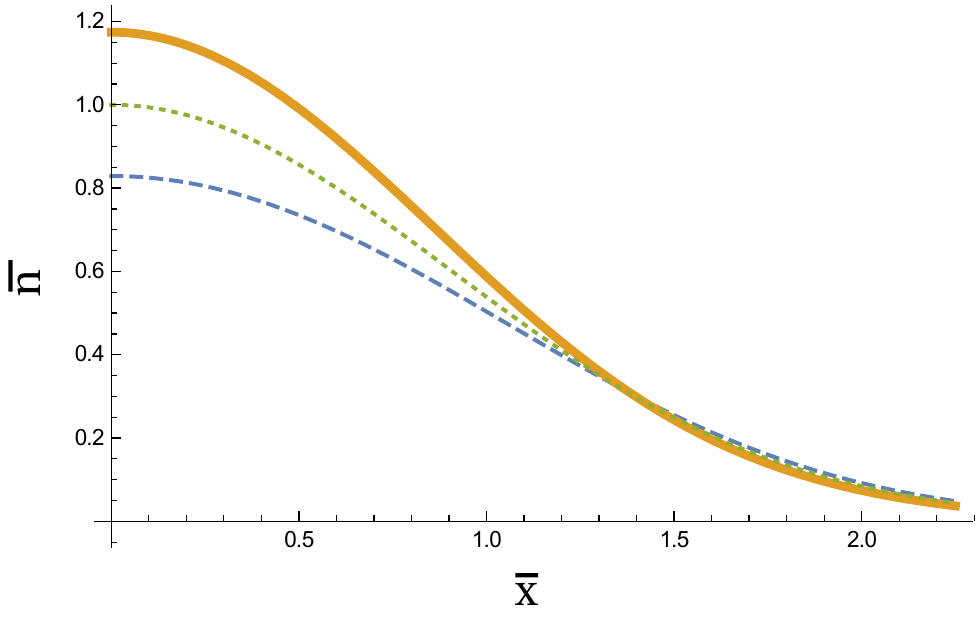}\ec
\bc\includegraphics[width=7.5cm]{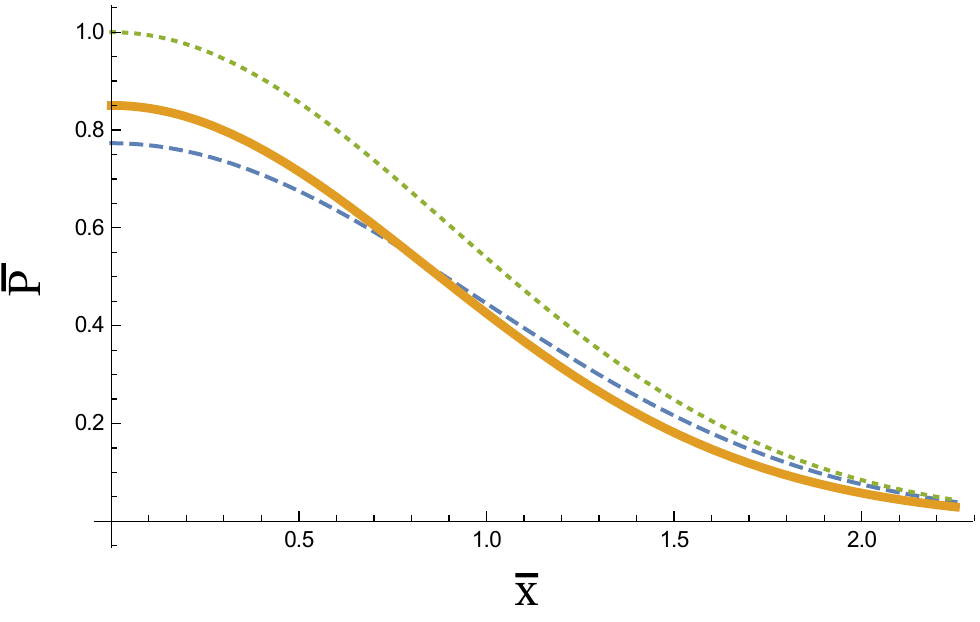}\ec
\caption{\label{fig_ninit}
Initial density $\bar{n}(\bar{x})$ and pressure $\bar{P}(\bar{x})$ of 
a trapped gas at $E/(NE_F)=1.49$. 
The full line shows the result for the MIT equation of state, the dotted line
is the Boltzmann limit, and the dashed line is a free Fermi gas. The distance
is given in units of $\tilde{\omega}_x^{-1}x_0$, and the density is in units
of the central density of a free Boltzmann gas.}
\end{figure}

 The equation of state determines solutions of the equation of hydrostatic 
equilibrium, $\vec\nabla P =-n\vec\nabla V_{\it ext}$. The Gibbs-Duhem
relation $dP=nd\mu$ implies that solutions are of the form $n(x)=n(\mu(x),T)$ 
where $n(\mu,T)$ is the equilibrium density and $\mu(x)=\mu_0-V_{\it ext}(x)$. 
The density of the homogeneous gas is given by $n(\mu,T) = \lambda^{-3}
g(\zeta)$ with
\be 
\label{app_5}
g(\zeta) = -{\it Li}_{3/2}(-\zeta^{-1})h(\zeta)
 + \zeta {\it Li}_{5/2}(-\zeta^{-1})h'(\zeta)\, . 
\ee
In dimensionless variables the solution to the hydrostatic equation is
\bea
\bar{n}(\bar{x}) &=& 3\left(\frac{T}{T_F}\right)^3 
      g\left(\zeta_0\exp\left( 
         \frac{\tilde{\omega}_i^2\bar{x}_i^2}{2\bar{T}}\right)\right)\,  , \\
\bar{P}(\bar{x}) &=& \frac{9}{2}\left(\frac{T}{T_F}\right)^4 
      f\left(\zeta_0\exp\left( 
         \frac{\tilde{\omega}_i^2\bar{x}_i^2}{2\bar{T}}\right)\right)\, , 
\eea
where $\bar{T}=3T/(2T_F)$. For a given $T/T_F$ the inverse fugacity $\zeta_0$ 
at the center of the trap is determined by the condition
\be 
\label{trap_zeta0}
 \frac{3}{(2\pi)^{3/2}}\left(\frac{T}{T_F}\right)^3 
\int d^3x\, g\left( \zeta_0 \exp\left( \frac{x^2}{2}\right)\right) 
 \equiv 1 \, . 
\ee
This equation has to be solved numerically. In the high temperature 
limit $\zeta_0= 6(T/T_F)^3$. The experiments do not measure $T/T_F$,
but only the total energy $E/(NE_F)$. We can compute the energy using
the virial theorem. We get
\be 
\label{trap_EoEF}
 \frac{E}{NE_F} = \left(\frac{T}{T_F}\right)
  \frac{\int d^3x\, x^2 g\left( \zeta_0 \exp\left( \frac{x^2}{2}\right)\right)}
 {\int d^3x\, g\left( \zeta_0 \exp\left( \frac{x^2}{2}\right)\right)}\,  . 
\ee
In the high temperature limit $E/(NE_F)=3(T/T_F)$. We can use 
equ.~(\ref{trap_zeta0}) and (\ref{trap_EoEF}) to compute both $T/T_F$ and 
$\zeta$ at the trap center as a function of $E/(NE_F)$. Note that these 
relations do not explicitly depend on $N$, but the units of distance, 
density, and pressure implicitly depend on the number of particles. 
Figure~\ref{fig_EoEf} shows $E/(NE_F)$ as a function of $T/T_F$. We observe 
that even though the full energy density and pressure differ significantly 
from that of a free gas, the relation between $E$ and $T$ of a trapped 
interacting gas is close to that of a trapped free gas, except at very 
low temperature. As an example we consider the case $E/(NE_F)=1.49$. We find
\be
 \left(\frac{T}{T_F}\right)_{\it full} = 0.538,\hspace{0.3cm}
 \left(\frac{T}{T_F}\right)_{\it free} = 0.496,\hspace{0.3cm}
 \zeta_0 = 1.154.
\ee
The corresponding density and pressure profiles are shown in 
Fig.~\ref{fig_ninit}. We compare to the density and pressure of 
a free Boltzmann and Fermi gas at the same $E/(NE_F)$.

\begin{figure}[t]
\bc\includegraphics[width=7.5cm]{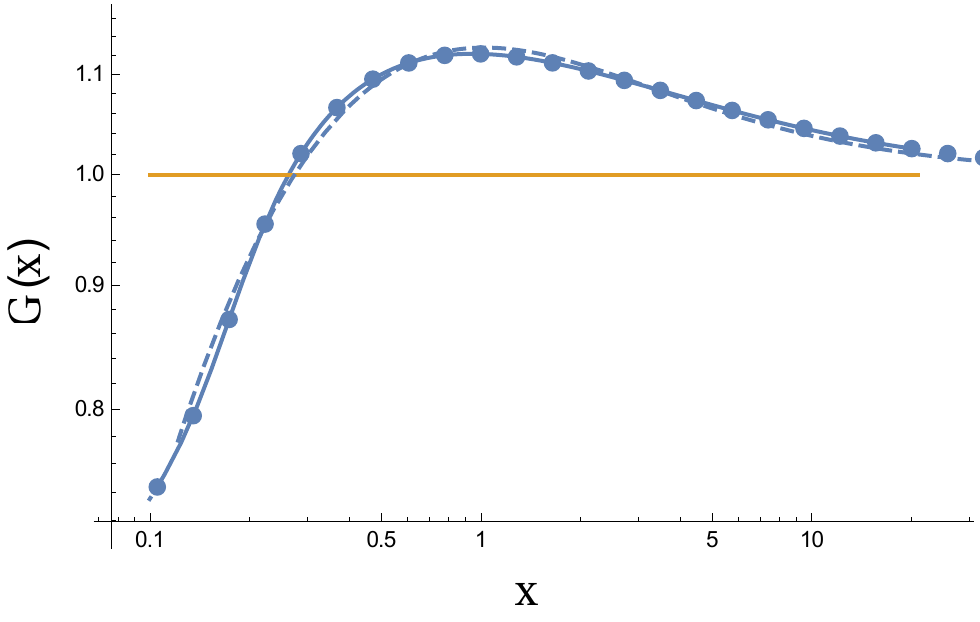}\ec
\caption{\label{fig_Tcor}
Temperature correction factor $G(x)$. The points and the solid 
line show the function $G(x)$ computed from the parametrization
of $h(\zeta)$ given in the text. The dashed line is the fit in 
equ.~(\ref{G_fit}).}
\end{figure}

 Below we will compare the hydrodynamic evolution initialized with the 
free Boltzmann gas to that initialized with the correct density and pressure
profile of an interacting gas. One complication is that for a free Boltzmann 
gas the temperature is given by $P/n$, whereas for the full equation of state 
the temperature is a more complicated function of $P$ and $n$. In particular, 
for the profiles shown in Fig.~\ref{fig_ninit} the ratio $P/n$ is not  
spatially constant.

\begin{figure}[t]
\bc\includegraphics[width=7.5cm]{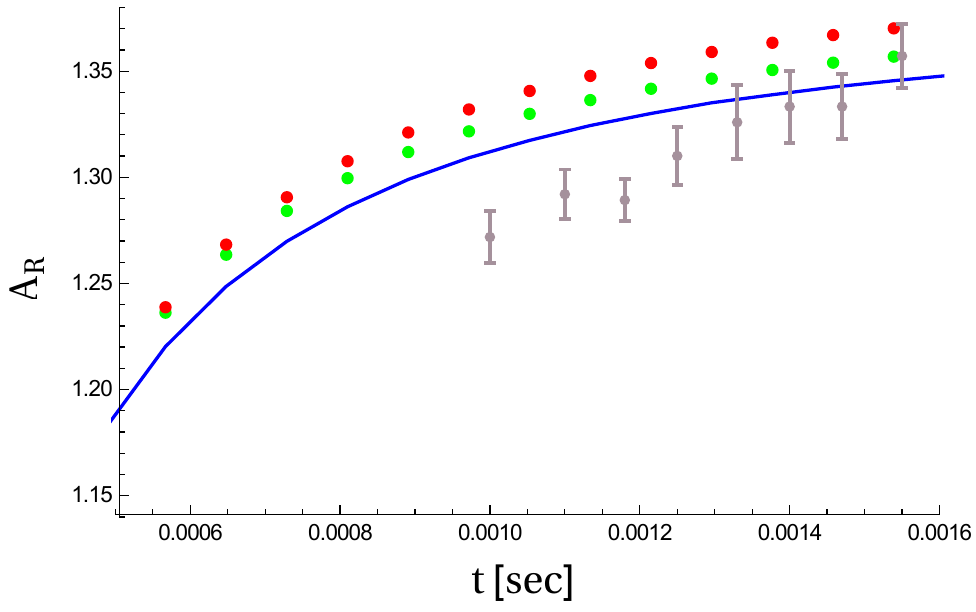}\ec
\caption{\label{fig_TOF_EOS}
This figure shows the time evolution of the aspect ratio $\sigma_x/
\sigma_y$ (extracted from a Gaussian fit) for $E/(NE_F)=1.49$. The green
dots show a hydrodynamic simulation with Gaussian (free Boltzmann gas)
initial conditions, and the red dots are obtained using the initial
conditions shown in Fig.~\ref{fig_ninit}. The solid line takes into
account the exact equation of state in computing the temperature 
dependence of the shear viscosity. The gray data points are measurements
for $E/(NE_F)=1.49$ from Joseph et al.}
\end{figure}

 In order to determine the temperature we follow the method suggested
in Appendix A of \cite{Schafer:2010dv}. Consider the function
\be 
 F(\zeta) = \frac{2f(\zeta)^{3/2}}{g(\zeta)^{5/2}}.
\ee
From equ.~(\ref{app_1}) and (\ref{app_5}) we know that this function
is proportional to $P^{3/2}/n^{5/2}$. The fugacity can then be computed 
using the inverse function $F^{-1}(x)$. We have
\be
\zeta = F^{-1}\left( \frac{2}{(2\pi)^{3/2}}\frac{(mP)^{3/2}}{n^{5/2}}\right).
\ee
From the fugacity we can compute the temperature
\be 
 T = \frac{g(\zeta)}{f(\zeta)}\, \frac{P}{n}\, .
\ee
In dimensionless units, this result can be written as 
\be 
\bar{T} = G(x) \frac{\bar{P}}{\bar{n}}\, , 
\ee
where we have defined the correction factor 
\be 
 G(x) = \frac{g(F^{-1}(x))}{f(F^{-1}(x))}\, , 
\ee
and the variable  
\be 
 x = \frac{16}{9}\left(\frac{3}{2}\frac{T}{T_F}\right)^{3/2} 
     \frac{\bar{P}^{3/2}}{\bar{n}^{5/2}}\, . 
\ee
Here $T/T_F$ is the initial temperature of the trap. This quantity
enters through the overall normalization of the density. The 
function $G(x)$ is fixed once $h(\zeta)$ has been determined. However,
since the definition of $G(x)$ involves inverse functions it has 
to be computed numerically. In practice, we have parametrized 
$G(x)$ in analogy with $h(\zeta)$, 
\be 
\label{G_fit}
 G(x) = \frac{1+d_1/x+d_2/x^2}{1+d_3/x+d_4/x^2}\, . 
\ee
We find
\bea 
d_1 = 1.8052, &\hspace{0.3cm}&
d_2 =-0.0022, \\
d_3 = 1.3668, &\hspace{0.3cm}&
d_4 = 0.1179. \nonumber 
\eea
The function $G(x)$ together with the fit in equ.~(\ref{G_fit})
is shown in Fig.~\ref{fig_Tcor}. Note that this parameterization
is restricted to the regime $T\gsim T_c$. 

 Finally, we can study the effect of the equation of state on
the anisotropic expansion. Figure~\ref{fig_TOF_EOS} shows the case
$E/(NE_F)=1.49$. The green dots show the aspect ratio computed using
initial conditions obtained from the free equation of state, and the 
red dots show the result for the initial conditions shown in 
Fig.~\ref{fig_ninit}. The red dots are computed using $\eta\sim 
T^{3/2}$ with $T=P/n$, and the solid line is computed using the 
full equation of state, including the correction factor $G(x)$. For 
comparison, we also show a set of 
aspect ratio measurements from \cite{Joseph:2014}. No attempt was
made to fit the data by adjusting $\eta_0$. We observe that the 
initial state has a significant effect on the time evolution. This
effect is more than compensated by the effect of using the correct
temperature $T=T(P,n)$.

\end{appendix}



\begin{thebibliography}{20}

\bibitem{Joseph:2014}
J.~A.~Joseph, E.~Elliott, J.~E.~Thomas,
``Shear viscosity of a universal Fermi gas near the superfluid phase 
transition,''
Phys.\ Rev.\ Lett.\ {\bf 115}, 020401 (2015) 
[arXiv:1410.4835 [cond-mat.quant-gas]].

\bibitem{Ku:2011}
M.~J.~H.~Ku, A.~T.~Sommer, L.~W.~Cheuk, and M.~W.~Zwierlein,
``Revealing the Superfluid Lambda Transition in the Universal
Thermodynamics of a Unitary Fermi Gas,''
Science 335, 563 (2012)
[arXiv:1110.3309 [cond-mat.quant-gas]].

\bibitem{Schafer:2009dj} 
T.~Sch\"afer and D.~Teaney,
``Nearly Perfect Fluidity: From Cold Atomic Gases to Hot Quark Gluon Plasmas,''
Rept.\ Prog.\ Phys.\  {\bf 72}, 126001 (2009)
[arXiv:0904.3107 [hep-ph]].

\bibitem{Adams:2012th} 
A.~Adams, L.~D.~Carr, T.~Sch\"afer, P.~Steinberg and J.~E.~Thomas,
``Strongly Correlated Quantum Fluids: Ultracold Quantum Gases, Quantum 
Chromodynamic Plasmas, and Holographic Duality,''
New J.\ Phys.\  {\bf 14}, 115009 (2012)
[arXiv:1205.5180 [hep-th]].

\bibitem{Schaefer:2014awa} 
T.~Sch\"afer,
``Fluid Dynamics and Viscosity in Strongly Correlated Fluids,''
Ann.\ Rev.\ Nucl.\ Part.\ Sci.\  {\bf 64}, 125 (2014)
[arXiv:1403.0653 [hep-ph]].

\bibitem{Kovtun:2004de} 
P.~Kovtun, D.~T.~Son and A.~O.~Starinets,
``Viscosity in strongly interacting quantum field theories from black hole 
physics,''
Phys.\ Rev.\ Lett.\  {\bf 94}, 111601 (2005)
[hep-th/0405231].

\bibitem{Bloch:2007}
I.~Bloch, J.~Dalibard, W.~Zwerger,
``Many-Body Physics with Ultracold Gases''
Rev.\ Mod.\ Phys.\ {\bf 80}, 885 (2008)
[arXiv:0704.3011].

\bibitem{Giorgini:2008}
S.~Giorgini, L.~P.~Pitaevskii, S.~Stringari, 
``Theory of ultracold atomic Fermi gases''
Rev.\ Mod.\ Phys.\ {\bf 80} 1215 (2008)
[arXiv:0706.3360].

\bibitem{Kinast:2004b}
J.~Kinast, A.~Turlapov, J.~E.~Thomas,
``Breakdown of Hydrodynamics in the Radial Breathing Mode of a 
Strongly-Interacting Fermi Gas,''
Phys.\ Rev.\ A {\bf 70}, 051401(R) (2004)
[arXiv:cond-mat/0408634 [cond-mat.soft]].

\bibitem{Schafer:2007pr}
T.~Sch\"afer,
``The Shear Viscosity to Entropy Density Ratio of Trapped Fermions in the
Unitarity Limit,''
Phys.\ Rev.\  A {\bf 76}, 063618 (2007)
[arXiv:cond-mat/0701251].

\bibitem{Turlapov:2007}
A.~Turlapov, J.~Kinast, B.~Clancy, L.~Luo, J.~Joseph, J.~E.~Thomas,
``Is a Gas of Strongly Interacting Atomic Fermions a Nearly Perfect Fluid''
J.\ Low Temp.\ Phys.\ {\bf 150}, 567 (2008)
[arXiv:0707.2574].
 
\bibitem{Bruun:2007}
G.~M.~Bruun, H.~Smith,
``Frequency and damping of the Scissors Mode of a Fermi gas'',
Phys. Rev. A {\bf 76}, 045602 (2007) 
[arXiv:0709.1617].

\bibitem{Cao:2010wa}
C.~Cao, E.~Elliott, J.~Joseph, H.~Wu, J.~Petricka, T.~Sch\"afer
and J.~E.~Thomas,
``Universal Quantum Viscosity in a Unitary Fermi Gas,''
Science {331}, 58 (2011)
[arXiv:1007.2625 [cond-mat.quant-gas]].

\bibitem{Elliott:2013b}
E.~Elliott, J.~A.~Joseph, J.~E.~Thomas,
``Anomalous minimum in the shear viscosity of a Fermi gas,''
Phys.\ Rev.\ Lett.\  {\bf 113}, 020406 (2014)
[arXiv:1311.2049 [cond-mat.quant-gas]].

\bibitem{Elliott:2013}
E.~Elliott, J.~A.~Joseph, J.~E.~Thomas,
``Observation of conformal symmetry breaking and scale invariance in 
expanding Fermi gases,''
Phys.\ Rev.\ Lett.\ {\bf 112}, 040405 (2014)
[arXiv:1308.3162 [cond-mat.quant-gas]].

\bibitem{Brewer:2015hua} 
J.~Brewer, M.~Mendoza, R.~E.~Young and P.~Romatschke,
``Lattice Boltzmann simulations of a strongly interacting two-dimensional 
Fermi gas,''
Phys.\ Rev.\ A {\bf 93}, no. 1, 013618 (2016)
[arXiv:1507.05975 [cond-mat.quant-gas]].

\bibitem{Ackermann:2000tr} 
K.~H.~Ackermann {\it et al.} [STAR Collaboration],
``Elliptic flow in Au + Au collisions at $(s_{NN})^{1/2}$ = 130 GeV,''
Phys.\ Rev.\ Lett.\  {\bf 86}, 402 (2001)
[nucl-ex/0009011].

\bibitem{Adler:2003kt} 
S.~S.~Adler {\it et al.} [PHENIX Collaboration],
``Elliptic flow of identified hadrons in Au+Au collisions at $(s_{NN})^{1/2}$
= 200-GeV,''
Phys.\ Rev.\ Lett.\  {\bf 91}, 182301 (2003)
[nucl-ex/0305013].

\bibitem{Aamodt:2010pa} 
K.~Aamodt {\it et al.} [ALICE Collaboration],
``Elliptic flow of charged particles in Pb-Pb collisions at 2.76 TeV,''
Phys.\ Rev.\ Lett.\  {\bf 105}, 252302 (2010)
[arXiv:1011.3914 [nucl-ex]].

\bibitem{Bluhm:2015raa} 
M.~Bluhm and T.~Sch\"afer,
``Dissipative fluid dynamics for the dilute Fermi gas at unitarity: 
Anisotropic fluid dynamics,''
Phys.\ Rev.\ A {\bf 92}, no. 4, 043602 (2015)
[arXiv:1505.00846 [cond-mat.quant-gas]].

\bibitem{Florkowski:2010} 
W.~Florkowski and R.~Ryblewski,
``Highly-anisotropic and strongly-dissipative hydrodynamics for 
early stages of relativistic heavy-ion collisions,''
Phys.\ Rev.\ C {\bf 83}, 034907 (2011)
[arXiv:1007.0130 [nucl-th]].

\bibitem{Martinez:2010sc} 
M.~Martinez and M.~Strickland,
``Dissipative Dynamics of Highly Anisotropic Systems,''
Nucl.\ Phys.\ A {\bf 848}, 183 (2010)
[arXiv:1007.0889 [nucl-th]].

\bibitem{Pantel:2014jfa} 
P.~A.~Pantel, D.~Davesne and M.~Urban,
``Numerical solution of the Boltzmann equation for trapped Fermi gases with 
in-medium effects,''
Phys.\ Rev.\ A {\bf 91}, 013627 (2015)
[arXiv:1412.3641 [cond-mat.quant-gas]].

\bibitem{Bluhm:2015bzi} 
M.~Bluhm and T.~Sch\"afer,
``Model-independent determination of the shear viscosity of a trapped 
unitary Fermi gas: Application to high temperature data,''
Phys.\ Rev.\ Lett.\  {\bf 116}, no. 11, 115301 (2016)
[arXiv:1512.00862 [cond-mat.quant-gas]].

\bibitem{Bruun:2005}
G.~M.~Bruun, H.~Smith,
``Viscosity and thermal relaxation for a resonantly interacting 
Fermi gas'',
Phys.\ Rev.\ A {\bf 72}, 043605 (2005) 
[cond-mat/0504734].

\bibitem{Schafer:2010dv} 
T.~Sch\"afer,
``Dissipative fluid dynamics for the dilute Fermi gas at unitarity: Free 
expansion and rotation,''
Phys.\ Rev.\ A {\bf 82}, 063629 (2010)
[arXiv:1008.3876 [cond-mat.quant-gas]].

\bibitem{Colella:1984}
P.~Colella, P.~R.~Woodward, 
``The Piecewise Parabolic Method (PPM) for Gas-Dynamical Simulations,''
J.\ Comp.\ Phys. {\bf 54}, 174 (1984).

\bibitem{Blondin:1993}
J.~M.~Blondin, E.~A.~Lufkin,
``The piecewise-parabolic method in curvilinear coordinates,''
Astrophys.\ J.\ Supp.\ Ser.\ {\bf 88}, 589 (1993).

\bibitem{Enss:2010qh} 
T.~Enss, R.~Haussmann and W.~Zwerger,
``Viscosity and scale invariance in the unitary Fermi gas,''
Annals Phys.\  {\bf 326}, 770 (2011)
[arXiv:1008.0007 [cond-mat.quant-gas]].

\bibitem{Wlazlowski:2013owa} 
G.~Wlazlowski, P.~Magierski, A.~Bulgac and K.~J.~Roche,
``The temperature evolution of the shear viscosity in a unitary Fermi gas,''
Phys.\ Rev.\ A {\bf 88}, 013639 (2013)
[arXiv:1304.2283 [cond-mat.quant-gas]].

\bibitem{Rupak:2007vp} 
G.~Rupak and T.~Sch\"afer,
``Shear viscosity of a superfluid Fermi gas in the unitarity limit,''
Phys.\ Rev.\ A {\bf 76}, 053607 (2007)
[arXiv:0707.1520 [cond-mat.other]].

\bibitem{Zwerger:2016xma} 
W.~Zwerger,
``Strongly Interacting Fermi Gases,''
Proceedings of the International School of Physics "Enrico Fermi"
- Course 191 "Quantum Matter at Ultralow Temperatures" edited by M.~Inguscio,
W.~Ketterle, S.~Stringari and G.~Roati (2016), pp. 63-142
[arXiv:1608.00457 [cond-mat.quant-gas]].

\bibitem{Nascimbene:2009}
S.~Nascimbene, N.~Navon, K.~Jiang, F.~Chevy, C.~Salomon, 
Nature 463, 1057 (2010)
[arXiv:0911.0747[cond-mat.quant-gas]].

\bibitem{Romatschke:2012sf} 
P.~Romatschke and R.~E.~Young,
``Implications of hydrodynamic fluctuations for the minimum shear viscosity 
of the dilute Fermi gas at unitarity,''
Phys.\ Rev.\ A {\bf 87}, no. 5, 053606 (2013)
[arXiv:1209.1604 [cond-mat.quant-gas]].
\end{thebibliography}
\end{document}